\documentclass[preprint,amsmath,amssymb]{revtex4}
\usepackage{graphicx}
\usepackage{dcolumn}
\usepackage{bm}
\begin{document}
\title{Optical Aharonov-Bohm effect due to toroidal moment inspired by general relativity} 
%\date{\today}
\author{${\rm A. \;Besharat}^{\;(a)}$ 
\footnote {Electronic
address:~afshin.besharat@ut.ac.ir}, 
${\rm M.\;Miri}^{\;(a)}$\footnote{Electronic
address:~mirfaez\_miri@ut.ac.ir}
and ${\rm M.\;Nouri}$-${\rm Zonoz}^{\;(a,b)}$\footnote {Electronic
address:~nouri@ut.ac.ir \; (Corresponding author)}}
\affiliation{(a): Department of Physics, University of Tehran, North Karegar Ave., Tehran 14395-547, Iran.\\ 
(b): School of Astronomy, Institute for Research in Fundamental Sciences, P O Box 19395-5531, Tehran, Iran}
\begin{abstract}
We study the analogy between propagation of light rays in a stationary curved spacetime and in a toroidal (meta-)material. After introducing a novel 
gravitational analog of the index of refraction of a magneto-electric medium, it is argued that light rays not only feel a Lorentz-like force in a magneto-electric 
medium due to the non-vanishing curl of the toroidal moment, but also there exists an optical analog of Aharonov-Bohm effect for the rays traveling 
in a region with a curl-free toroidal moment. Experimental realization of this effect could utilize either a multiferroic material or a toroidal metamaterial.
\end{abstract}
\maketitle
 
\section{Introduction}
Apart from yet being the most successful theory of gravity, general relativity and the exact solutions of Einstein field equations, in recent years have turned into a 
very powerful tool to study linear and nonlinear phenomena in other branches of physics. An interesting 
context, in which general relativity has already proved amazingly fruitful in this regard, was formed around the analogy between a curved background and a material 
medium with respect to light propagation.\\
The idea could be traced back the study of electromagnetism and light propagation in curved spacetimes and the observation that the underlying spacetime 
could be taken as a medium assigned with an index of refraction \cite{LL}. 
The same interpretation was reinforced by noting that Maxwell's equations in a curved spacetime lead to constitutive equations with magneto-electric coupling, 
in which electric and magnetic fields and their corresponding features are intertwined through the geometry of the underlying spacetime.
In this way, one could establish a correspondence between geometric entities assigned to a curved spacetime and
the electromagnetic features of a  medium such as electric permittivity and magnetic permeability  \cite{LL,pleban}.\\
Meanwhile studies, both theoretical and experimental, on metamaterials have flourished with the beginning of the 21st century, mainly due to the advancements 
in small-scale technologies, including nanotechnolgy. 
Metamaterials are artificially designed matter which have very  interesting electromagnetic properties, 
not found in simple or composite matter in nature. Their nontrivial properties, such as negative refractive index \cite{Smith}, originate not from 
the composition of their constituents, but from the way these subwavelength macroscopic constituents are assembled.\\ 
Another key development was the introduction of {\it transformation optics} \cite{Pendry}, which formulates how electromagnetic field lines and the 
corresponding light rays 
could be redirected by deforming the underlying space in which they
are embedded, very much like rerouting a river by deforming its bed. 
This process could be looked upon as a coordinate transformation from the initial configuration of the electromagnetic fields
in a Cartesian coordinate to their final configuration in a distorted coordinate system, justifying the nickname 
{\it optical conformal transformation or mapping} \cite{Leon, Leon1}.\\
Putting the above  elements together, it was proposed that one could control the light propagation by a carefully designed metamaterial through the application 
of the above transformation. 
This simple idea is the key idea behind employing metamaterials to build cloaking devices in which, light is directed by a carefully assembled metamaterial to 
avoid a designated region of space \cite{Pendry,Leon}.\\
On the other hand, the geometric nature of general relativity, introducing gravity not as a force but as the geometry of spacetime, leads to a machinery, based on Riemannian geometry  
which allows one to calculate paths of massive and massless particles, in particular light paths, in a given spacetime. Hence, for a given spacetime, regardless of the 
matter distribution producing it, one is able in principle to calculate its null geodesics and  the metamaterial analog of this spacetime, designed 
with the equivalent electromagnetic specifications, will reproduce the same optical paths. \\
Obviously the more exotic a spacetime may look, the more interesting is expected to be the behavior of its light trajectories and that is exactly the case with light paths 
in the geometry of black holes.
Consequently in the metamaterial analogs of black holes, one could duplicate the same light paths, however strange they may seem.
For example in the simplest spherically symmetric black hole, namely the Schwarzschild black hole, there is a {\it photon sphere} which is formed by  unstable circular photon 
orbits at a radius $1.5$ times 
that of the black hole's event horizon \cite{Virb}. Interestingly enough, it is found, through simulations, that the metamaterial analog of this black hole exhibits the same 
photon sphere property \cite{Genov, Chen}. Later experimental demonstration of gravitational lensing effects near the photon sphere was carried out in \cite{Sheng}.
Metamaterial designs are suggested that could form an {\it optical black hole} in the sense that it is a complete light absorber, and are 
hoped to find application in many areas \cite{Nariman}. Optical analogs of wormholes 
were also suggested by designing metamaterial according to the electromagnetic parameters mimicking that of a wormhole solution of Einstein field equations, 
connecting two distant regions in 
a spacetime continuum \cite{Green}. Even it is suggested that one could mimic {\it Hawking radiation} in optical analogs of a black hole  \cite{Leon2,Linder}. \\
In the present article,  motivated by the gravitational Aharonov-Bohm effect, first we use the $1+3$ (or threading) formulation of
spacetime decomposition to exhibit an interesting analogy between electromagnetic wave propagation in a stationary curved spcetime and in a magneto-electric (meta-)material, 
and then employ this analogy to show that there should be an optical analog of this effect in (meta-)materials with toroidal moment. 
The magneto-electric coupling in (meta-)materials mimicking stationary spacetimes, are constrained to those due to the toroidal moment
originating from the antisymmetric part of the linear magneto-electric tensor. 
A natural and direct consequence of this analogy is the existence of a Lorentz-like force acting on light, similar to
what we have for charged particles moving in a magnetic field.  Also employing the same analogy, we predict an optical analog of the Aharonov-Bohm effect in a 
toroidal (meta-)material. This effect is different from the dynamic version of the usual Aharonov-Bohm effect which is expected to affect
beams of {\it charged} particles in the presence of non-radiating sources formed by a special combination of interfering toroidal and electric dipoles \cite{Bash}. \\
The outline of the paper is as follows. In the next section, in the context of gravitoelectromagnetism, we introduce the idea of a spacetime as a medium assigned 
with an index of refraction with respect to 
the light trajectories. For a {\it stationary} spacetime we arrive at a novel form for the spacetime index of refraction, introduced here for the first time in the literature,
which enables us to interpret it as the gravitational analog of the magneto-electric index of refraction in a material with toroidal moment.
In section three this analogy is extended to Maxwell equations in a stationary spacetime and their corresponding constitutive equations which include  magneto-electric 
terms. In section four we discuss in more detail stationary spacetime metrics and their magneto-electric (Meta-)material analogs with respect to their corresponding 
magneto-electric tensor.\\ 
In section five after a brief introduction of the gravitational Aharonov-Bohm effect we discuss our prediction of the optical analog of the Aharonov-Bohm effect 
in toroidal (meta-)materials with any value for the toroidal moment and show that it reduces to the previous result in the limit of small toroidal moment.
Finally we discuss and summarize our results in the last section.\\
Throughout we use the (+,-,-,-) signature for spacetime metric and our convention for indices is such that the Latin indices run from 1 to 3 while the Greek ones 
run from 0 to 3.
%%%%%%%%%%%%%%%%%%%%%%%%%%%%%%%%%%%%%%%%%%%%%%%%%%%%%%%%%%%%%%%%%%%%%%%%%%%%%%%%%%%%%%%5
\section{curved Spacetime as a medium}
Any physical realization of 3-dimensional quantities in a curved background requires a decomposition of spacetime into spatial and temporal sections.
There are two main decomposition formalisms called $1+3$  and $3+1$ or alternatively {\it threading} and {\it foliation} formalisms respectively 
\footnote{For a comparison between the two formalisms refer to \cite{Ghare}.}. To motivate 
the idea of a curved spacetime as a medium, we start with introducing the threading formalism which will naturally lead to two interesting
analogies between  a curved spacetime and  a medium with respect to light propagation and electromagnetic constitutive equations in either of them.\\
In the threading formulation of spacetime decomposition, the spatial and temporal distances between two infinitesimally close fundamental observers are obtained by
sending and receiving light signals  on their worldlines. The outcome is the following  decomposed spacetime metric, 
\begin{equation}\label{ds0}
d{s^2} = d\tau_{syn}^2 - d{{l}^2} = {g_{00}}{(d{x^0} - {{A_g}_i }d{x^i })^2} - {{\gamma}_{i j }}d{x^i }d{x^j }
\end{equation}
where ${{A_g}_i } \equiv  - \frac{{{g_{0i }}}}{{{g_{00}}}}$ and
\begin{equation}\label{gamma0}
{{\gamma} _{ij}} =  - {g_{i j }} + \frac{{{g_{0i }}{g_{0j }}}}{{{g_{00}}}}, 
\end{equation}
is the spatial metric of a 3-space called a {\it quotient space/manifold}, on which $d{{l}}$ gives the element of spatial distance
between any two nearby events. 
Also the so called {\it synchronized proper time} 
between any two events is given by $d{\tau _{syn}} = \sqrt {{g_{00}}} (d{x^0} - {{A_g}_i }d{x^i })$, so that any two simultaneous events 
have a coordinate-time difference of $d x^0 = {{A_g}_i } dx^i$.
The spatial three-metric is invariant under the special coordinate transformation defined 
by $\frac{{\partial}{\acute{x^{i}}}}{{\partial}{x^{0}}}=0$ and $\frac{{\partial}{\acute{x^{0}}}}{{\partial}{x^{0}}}=1$.
In addition, we may define the three-tensors, vectors, and scalars under the same coordinate transformation. Specifically, $g_{00}$ and $g^{00}$ are 
three-scalars while  ${A_g}_{i} \equiv -\frac{g_{0i}}{g_{00}}$ 
are the components of a covariant 3-vector called {\it gravitomagnetic potential} (see Eq. (5) below), and $\gamma^{ij}=-g^{ij}$ is the inverse of the 3-metric. We can consistently raise or lower three 
dimensional indexes using the spatial three-metric and its inverse. Moreover, the curl of a three-vector is defined using the three-dimensional  
anti-symmetric tensor, $\eta_{ijk}=\sqrt{\gamma}e_{ijk}$ (and ${\eta^{ijk}}=\frac{e^{ijk}}{\sqrt{\gamma}}$), with ${\gamma}$, the 
determinant of the 3-metric and $e^{ijk}=e_{ijk}=1$, the Cartesian Levi-Civita symbol.\\
The {\it gravitational} 3-force acting on a test particle moving on a {\it spacetime} geodesic, in a stationary background is given by \cite{LL,lynden1998classical}
\begin{gather}\label{force}
\vec{f}=\frac{{\varepsilon}_{0}}{\sqrt{g_{00}}}[-\vec{\nabla}\ln(\sqrt{g_{00}})+\sqrt{g_{00}}\frac{\vec{v}}{c}\times(\vec{\nabla}\times\vec{A_g})]\\
\vec{f}_g =  \frac{{\varepsilon}_{0}}{\sqrt{g_{00}}}\left( \vec {E_g} + \frac{\vec{v}}{c}\times \sqrt{g_{00}} \vec {B_g}\right)
\end{gather}
in which $\vec{v}$ is the three-velocity of the test particle, ${\varepsilon}_{0} = - c \frac{\partial{S}}{\partial{x^{0}}}$ is its conserved  energy with action $S$, and  
the gravitoelectric and gravitomagnetic 3-fields 
are defined as follows \footnote{All the differential
operations in these relations are defined in the 3-space with metric $\gamma_{ij}$. Specifically the curl and divergence of a 3-vector ${\bf V}$ are given by
$({\bf \nabla}\times{\bf V})^i = \frac{1}{\sqrt{\gamma}}e^{ijk} \partial_j V_k$ and 
$({\bf \nabla}.{\bf V}) = \frac{1}{\sqrt{\gamma}}\partial_i (\sqrt{\gamma}V^i)$ respectively.}
\begin{equation}
\vec{B_g} = \vec{\nabla}\times\vec{A_g}\;\;   \;\; \;\;\; \;\;\;
\vec{E_g} = -{\vec {\nabla}} \ln \sqrt{g_{00}}.
\end{equation} 
In the case of null rays (photons), $\vec {v} = c\hat {\bf k}$ 
where $\hat{\bf k}$ is the unit vector along the direction of propagation.
Compared with electromagnetism, the first term in the right hand side of \eqref{force} could be interpreted as the gravitoelectric 
force due to the gravitoelectric potential $\phi_g =\ln\sqrt{g_{00}}$ while the second term is interpreted as the gravitomagnetic force due to the gravitomagnetic field 
$\vec{B_g}$  with $\vec{A_g}$ as the gravitomagnetic vector potential. In the same spirit ${\epsilon}_{0}$ could be interpreted as 
the gravitoelectric charge of the test particle \cite{lynden1998classical}.\\
Now starting with Fermat's principle $\delta \int k_\alpha dx^\alpha =0$, and using the threading formulation of spacetime decomposition \cite{LL}, 
one arrives at the following relation
\begin{equation}\label{IR}
\delta \int \left( \frac{dl}{\sqrt{g_{00}}} + {A_g}_i dx^i \right)=0
\end{equation}
in which $dl$ is the spatial line element along the light path. Restricting the above result to the case of static spacetimes by setting  ${A_g}_i=0$, one  
could obviously assign an index of refraction, $n_0 = \frac{1}{\sqrt{g_{00}}}$ to the underlying {\it static} spacetime in analogy with a material medium \cite{LL}. 
Interestingly enough, in what follows, we show that an index of refraction could also be assigned to stationary spacetimes. To this end we rewrite \eqref{IR} in the following way
\begin{equation}\label{IRS}
\delta \int  \left( \frac{1}{\sqrt{g_{00}}} + {A_g}_i \frac{dx^i}{dl} \right) dl \equiv \delta \int  n dl = 0
\end{equation}
from which the index of refraction could  be read as 
\begin{equation}\label{IRS1}
n = \frac{1}{\sqrt{g_{00}}} + {A_g}_i \frac{dx^i}{dl}
\end{equation}
Using the facts that $k^2 = \frac{{k_0}^2}{g_{00}} - \gamma_{ij}k^i k^j = 0$ and the wave vector $k^i$ is along the spatial displacement vector, $dx^i$, leads 
to $k^i = \frac{{k_0}}{\sqrt{g_{00}}} \frac{dx^i}{dl}$ and $|{\bf k}| = \frac{k_0}{{\sqrt{g_{00}}}}$, whereby the above relation could be written in the 
following illuminating simple form \footnote{It should be noted that the 
spatial contravariant components of the 4-vector $k^\alpha$ are taken as a 3-vector in the $\gamma$-space with the corresponding covariant 3-vector
given by $k_i = \gamma_{ij}k^j$.}
\begin{equation}\label{IRS2}
n = n_0 + {\bf {A}_g} . {\bf{\hat k}}
\end{equation}
On the other hand it has been shown, interestingly enough,  that there exists a Lorentz-like force acting on light rays in materials with toroidal moment corresponding to 
the optical analog of the Lorentz force acting on a charged particle in a magnetic field. This was realized in the optical magnetoelectric effect 
in multiferroic materials \cite{sawada2005optical}.
The optical magnetoelectric effect in a material with toroidal moment ${\bf T({\bf r})}$,  in the geometrical optics limit, is characterized  by the following 
space-dependent index of refraction 
\begin{equation}\label{OME}
n({\bf r}) = n_0 ({\bf r}) +\alpha {\bf T({\bf r})}.{\bf{\hat k}}
\end{equation}
Comparing the above results shows that \eqref{IRS2} is an interesting 
gravitational analog of the above magnetoelectric index of refraction (for $\alpha =1 $). 
In this analogy the above introduced gravitomagnetic potential $\vec{A_g}$ \cite{lynden1998classical}, plays the role of 
the toroidal moment and in the same way its presence accounts for the breaking of both time-reversal and parity symmetries. 
To the best of our knowledge, both the relation \eqref{IRS2}, and the above mentioned analogy between the two refractive indices (i.e \eqref{IRS2} and \eqref{OME}), 
appear here for the first time in the literature.
Now if we turn the analogy around, the  quasi-Lorentz force \eqref{force} acting on test particles or light rays in stationary spacetimes implies 
that there should exist a Lorentz-like force acting on light rays in a toroidal (meta-)material with the curl of the toroidal moment acting as a magnetic-type field. 
%%%%%%%%%%%%%%%%%%%%%%%%%%%%%%%%%%%%%%%%%%%%%%%%%%%%%%%%%%%5
\section{Maxwell equations in a curved background and the Magnetoelectric constitutive relations}
Apart from the novel analogy introduced in the last section, there is another well known clue hinting towards the same analogy between a curved 
spacetime and a so called {\it bianisotropic} medium in a flat background spacetime. 
This is achieved by looking at the 
equations of  electrodynamics in a curved background employing the 1+3 spacetime decomposition through the introduction of the following electromagnetic fields \cite{LL},
\begin{equation}\label{Fields}
E_{i}=F_{0i}  \;\;,\;\; D^{i}=-\sqrt{g_{00}}F^{0i}  \;\; , \;\; B^{i}=-\frac{1}{2\sqrt{\gamma}}e^{ijk}F_{jk}  \;\; , \;\; 
H_{i}=-\frac{\sqrt{\gamma}}{2}e_{ijk}F^{jk} 
\end{equation}
which coincide, in the absence of curvature, with their definitions in flat spacetime.
Using these definitions, Maxwell's equations in the absence of charges and currents in a stationary spacetime are
\begin{equation}\label{max1}
\vec{\nabla}.\vec{B}=0 \;\; , \;\; \vec{\nabla}\times\vec{E}=-\frac{1}{c}\frac{1}{\sqrt{\gamma}}\frac{\partial}{\partial{t}}(\sqrt{\gamma}\vec{B}) \;\; , \;\;
 \vec{\nabla}.\vec{D}=0 \;\; , \;\; \vec{\nabla}\times\vec{H}=\frac{1}{c}\frac{1}{\sqrt{\gamma}}\frac{\partial}{\partial{t}}(\sqrt{\gamma}\vec{D}).
 \end{equation}
Now, we rewrite the above decomposed, Maxwell equations in the following form which is formally equivalent to the Maxwell equations in a bianisotropic medium 
in a {\it flat background} \cite{pleban}, 
$$
\frac{1}{\sqrt{{\gamma^{0}}}}{\partial}_{i}[\sqrt{{\gamma}^{0}}(\sqrt{{\gamma}}\frac{B^{i}}{\sqrt{{\gamma}^{0}}})]=0,\;\;\;\;\;\;
\frac{1}{\sqrt{{\gamma^{0}}}}{\partial}_{i}[\sqrt{{\gamma}^{0}}(\sqrt{{\gamma}}\frac{D^{i}}{\sqrt{{\gamma}^{0}}})]=0,
$$
 $$
\frac{e^{ijk}{\partial_{j}}E_{k}}{\sqrt{{\gamma}^{0}}}=-\frac{1}{c}
\frac{{\partial}}{{\partial}t}
(\frac{\sqrt{\gamma}B^{i}}{\sqrt{{\gamma}^{0}}}), \;\;\;\;\;\; \frac{e^{ijk}{\partial_{j}}H_{k}}{\sqrt{{\gamma}^{0}}}=\frac{1}{c}
\frac{{\partial}}{{\partial}t}
(\frac{\sqrt{\gamma}D^{i}}{\sqrt{{\gamma}^{0}}})
 $$
where, ${\gamma}^{0}$ is the determinant of the spatial metric ${\gamma^0_{ij}}$ in flat space and
in the same orthogonal curvilinear coordinate system used to define ${\gamma}$ \footnote{We notice that the above definitions of $D^i$ and $B^i$, differ 
from those introduced in \cite{pleban} by a $\sqrt{\gamma}$ factor.} . 
Defining the new 3-vectors 
$\vec{b}=\sqrt{{\gamma}}\frac{\vec{B}}{\sqrt{{\gamma}^{0}}}$ and $\vec{d}=\sqrt{{\gamma}}\frac{\vec{D}}{\sqrt{{\gamma}^{0}}}$, the Maxwell equations 
are now given by
\begin{equation}
 \vec{{\nabla}_{0}}.\vec{b}=0, \;\;\;\;\;\; \vec{{\nabla}_{0}}\times\vec{E}=-\frac{1}{c}\frac{{\partial}}{{\partial}t}\vec{b},
 \;\;\;\;\;\; \vec{{\nabla}_{0}}.\vec{d}=0, 
 \;\;\;\;\;\; \vec{{\nabla}_{0}}\times\vec{H}=\frac{1}{c}\frac{{\partial}}{{\partial}t}
 \vec{d} 
 \end{equation}
in which the {\it curl}  and {\it divergence} (with $\vec{{\nabla}_{0}}$) are defined with respect to the orthogonal curvilinear coordinates in flat space
(For a detailed discussion on the relation between the background metric of a medium  and its analog spacetime metric refer to \cite{schuster2017effective}). 
Using the above definitions of the 3-vectors $\vec{b}$ and $\vec{d}$
and the fact that the fields introduced in \eqref{Fields} are not independent, one is led to the following constitutive equations \cite{LL}, 
 \begin{equation}\label{constraint}
 d^{i}=\frac{{\gamma}^{ij}\sqrt{\gamma}}{\sqrt{\gamma^{0}}\sqrt{g_{00}}}E_{j}+\frac{e^{ijk}{A_g}_k}{\sqrt{{\gamma}^{0}}}H_{j},\;\;\;\;   
 b^{i}=\frac{{\gamma}^{ij}\sqrt{\gamma}}{\sqrt{\gamma^{0}}\sqrt{g_{00}}}H_{j}-\frac{e^{ijk}{A_g}_k}{\sqrt{{\gamma}^{0}}}E_{j} 
 \end{equation}
Comparison of the above equations with the electromagnetic constitutive equations, which describe the behavior of matter
under the influence of electromagnetic fields, once again shows that the spacetime metric is formally playing 
the role of a medium. This analogy is in accordance with the assignment of an index of refraction to a stationary spacetime with respect 
to light propagation, discussed in the previous section. 
In other words solving Maxwell equations \eqref{max1} with respect to the above constitutive relations in the geometric optics limit should lead to
the null trajectories in the underlying spacetime.  
The above analogy establishes an {\it opto-geometric correspondence} between a spacetime and a material medium through which one 
is able to replicate the same light paths in a
material or metamaterial medium with equivalent properties (e.g electric and magnetic susceptibilities) read from the above spacetime constitutive relations. \\
To account for the other side of this correspondence, consider the matter interacting with the time-independent electromagnetic field to be in thermodynamic equilibrium. 
A systematic way to derive the general response of such a material to the electromagnetic field is 
through the following expansion of the free energy \cite{fiebig2005revival} in terms of the electromagnetic fields as well as  {\it spontaneous} polarization and magnetization, 
 $ {P^{i}}_{(s)}$ and ${M^{i}}_{(s)}$ as the secondary sources; 
$$F(\vec{E},\vec{H})=F_{0}-{P^{i}}_{(s)}{E_{i}}-{M^{i}}_{(s)}{H_{i}}$$
$$-\frac{\epsilon^{ij}{E_{i}}{E_{j}}}{8\pi}-\frac{\mu^{ij}{H_{i}}{H_{j}}}{8\pi}-\alpha^{ij}{E_{i}}{H_{j}}-{\rm Higher \;\; order \;\;terms} $$ 
in which $\epsilon^{ij}$ and $\mu^{ij}$ are the electric permittivity and the magnetic permeability respectively, 
and $\alpha^{ij}$ is the magneto-electric tensor of matter {\it inducing} either polarization by a magnetic field or magnetization by an electric field. 
Note that the higher order terms only come into play when
the fields acting on matter are strong enough to be comparable to the Coulomb potential in the atomic scale. 
Here we only consider the linear terms which are enough to mimic a medium's response to an electromagnetic field with that of a spacetimes' response to the same field.\\
The thermodynamic stability condition imposes the following constraint on the magneto-electric tensor \cite{fiebig2005revival}
 \begin{equation}
\alpha^{ij} \leq\frac{\sqrt{{\epsilon^{ii}\mu^{jj}}}}{4\pi}.
\end{equation}
We have the following general expansion for the electric displacement field and the magnetic intensity
 \begin{equation}\label{D1}
 D^{i}=-{4\pi}\frac{\partial{F}}{\partial {E}_i}=\epsilon^{ij}E_{j}+4\pi{\alpha}^{ij}H_{j}+4\pi{P^{i}}_{(s)},
 \end{equation}
 \begin{equation}\label{B1}
 B^{i}=-{4\pi}\frac{\partial{F}}{\partial {H}_i}=\mu^{ij}H_{j}-4\pi{\alpha}^{ij}E_{j}+4\pi{M^{i}}_{(s)}
 \end{equation}
Although the above equations are obtained for time-independent EM fields, they are also valid for low frequency time-dependent EM fields \cite{Landau2}.
%%%%%%%%%%%%%%%%%%%%%%%%%%%%%%%%%%%%%%%%%%%%%%%%%%%%%%%%5 
\section{Stationary Spacetimes and their Magneto-electric (Meta-)material analogs}
In the absence of spontaneous polarization and magnetization, comparing \eqref{D1} and \eqref{B1} with \eqref{constraint}, we conclude that the Maxwell
equations in a material medium  and in a curved spacetime are equivalent if the following correspondences are held; 
\begin{equation}\label{ana1}
\mu^{ij}=\epsilon^{ij} \sim \frac{{\gamma}^{ij}\sqrt{\gamma}}{\sqrt{\gamma^{0}}\sqrt{g_{00}}} =-\frac{g^{ij}\sqrt{g}}{\sqrt{\gamma^{0}}g_{00}}
\end{equation}
\begin{equation}\label{ana2}
4\pi{\alpha}^{ij} = - 4\pi{\alpha}^{ji} \sim \frac{e^{ijk}{A_g}_k}{\sqrt{{\gamma}^{0}}} = -\frac{e^{ijk}g_{0k}}{\sqrt{{\gamma}^{0}}g_{00}}
\end{equation}
We note, on passing, that the nonvanishing spontaneous polarization and magnetization could be taken as the electromagnetic analogs of the gravitoelectric and 
gravitomagnetic fields originating from the presence of the charges and currents in exact stationary solutions of Einstein-Maxwell equations such as  
the Kerr-Newmann spacetime. Here our study is restricted to the analog responses of a vacuum stationary spacetime and a medium  to a test 
electromagnetic field.\\
Also by the non-covariant nature of the spacetime decomposition, leading to Eqs. \eqref{constraint} and \eqref{ana2}, it is expected that the 
gravitational analog of the magneto-electric tensor depends, not only on the employed 
coordinates and different coordinate patches, but also on the definitions of the 3-dimensional electromagnetic fields \cite{Gibbons}. For a covariant approach
to transformation optics in linear dielectric materials refer to \cite{Thompson}. \\
The general form of the magneto-electric tensor of a (meta-)material medium is given by \cite{Spaldin},
\begin{equation}\label{alpha}
{\alpha}^{ij} = S^{ij}+e^{ijk}\tau_{k} + \gamma \delta^{ij}, 
\end{equation}
in which $S^{ij}$ is the symmetric traceless part of the magneto-electric tensor, $\vec{{\tau}}$ is a 3-vector dual of its antisymmetric part, 
and $\gamma$ is a pseudo-scalar representing the trace of the magneto-electric tensor. Therefore, comparing equations \eqref{ana2} and  \eqref{alpha}, 
one can conclude that the magneto-electric response of a time independent 
spacetime to an electromagnetic wave is the analog of the magneto-electric response of a material medium represented by the 3-vector $\vec{{\tau}}$. 
In other words  $\vec{{\tau}}$ of a material medium is the analog of the gravitomagnetic potential $\vec{A_{g}}$ of a stationary
spacetime. A non-vanishing $\vec{A_{g}}$ in stationary spacetimes is a manifestation of the breaking of the time reversal symmetry. This is the gravitational analog of 
breaking of the same symmetry in material media with a non-vanishing $\vec{{\tau}}$ such as in the case of multiferroic materials in 
which $\vec{{\tau}}$ is the so called toroidal moment.\\ 
In geometrical optics, light behaves as particles and so in this limit equation \eqref{force} should also be applicable to light, and therefore in the corresponding 
metamaterial analog, we expect the light ray to be affected  by an electric-like force as well as a 
magnetic-like force. The rational behind this idea is that the wave equations in a curved spacetime and in the corresponding metamaterial 
analog (in the same flat space coordinates) are equal and in 
the geometric approximation the same behavior for light paths is expected in the analog counterparts. Therefore, one can 
duplicate the null geodesics  of a stationary spacetime, as the light paths in the corresponding metamaterial and vice versa \cite{fernandez2016anisotropic}. 
So through the above analogy, one expects the existence of a
Lorentz-like force on light rays in a metamaterial designed with respect to \eqref{ana1} and \eqref{ana2} in which  
the rotation of  $\vec{{\tau}}$ of the (meta-)material, plays the role of the gravitomagnetic field. Obviously, now the gravitomagnetic potential $A_g$ is 
the gravitational analog of the vector $\vec{{\tau}}$. Extending the above analogy to other gravitational effects and inspired by the  gravitational Aharonov-Bohm effect, 
in the next section, we discuss the {\it optical Aharonov-Bohm} effect which 
could be realized in metamaterials designed with the specific optical characteristics determined from  \eqref{ana1} and \eqref{ana2}.
\section{Optical Aharonov-Bohm Effect in toroidal (meta-)material}
Before discussing the optical Aharonov-Bohm effect we digress to briefly discuss the gravitational Aharonov-Bohm effect. Different versions of gravitational 
analogue of the Aharonov-Bohm effect are discussed in the literature all of which corresponding to a test particle moving in a
region of space in which either the gravitomagnetic field ($\vec{B_g}$) or the gravitoelectric field ($\vec{E_g}$) are  absent, but the particle is 
affected by the fluxes of the same fields in the regions of spacetime from which it is excluded \cite{Ford}. 
These two versions could be called gravitomagnetoic and gravitoelectric Aharonov-Bohm effects respectively \cite {Nouri-2013}. Obviously the 
gravitomagnetic Aharonov-Bohm effect is more analogous, specially in its form, to its electromagnetic counterpart which involves magnetic field 
and its potential. In the usual Aharonov-Bohm effect, 
for a time-independent 4-vector potential $A_\mu = (\Phi, A_i)$, 
a particle is influenced locally by an electric field ${\vec E} = -{\vec \nabla} \Phi({\bf r})$ and globally by the curl-free vector potential $A_i$ through 
its integral over a non-trivial closed curve, i.e
\begin{eqnarray}\label{int2}
\oint A_\mu dx^\mu = -\oint {A}_i dx^i \neq 0,
\end{eqnarray}
where the line integral is taken over a closed path in the region where ${\vec B} = {\vec \nabla}\times {\vec A} = 0$.\\
Employing the spacetime decomposition introduced in the last section,  the existence of the 
gravitational analog of \eqref{int2} is given by
\begin{eqnarray}\label{g00}
\oint_C {A_g}_i dx^i \neq 0.
\end{eqnarray}
As an example of the gravitational analog of the Aharonov-Bohm effect, it is shown in \cite{stachel1982globally} that the {\it interference of light} passing each 
side of a rotating dust cylinder, represented by the Van Stockum solution \cite{exact} of Einstein field equations, is affected by the gravitomagnetic potential of the cylinder. 
The effect is nothing but a phase change proportional 
to the above factor calculated along the light path in a region where there are no local effects due to rotation \footnote{A more faithful gravitational analog of the 
Aharonov-Bohm effect for a truly confined gravitomagnetic field could be illustrated for the toroidal metric of a
toroidal shell with rolling motions around its small cross section \cite{Donald3}.}.\\
Since the gravitational Aharonov-Bohm effect only depends on the condition \eqref{g00}, its optical counterpart is also expected
to depend only on the toroidal moment and be independent of the structure of  either $\epsilon^{ij}$ or $\mu^{ij}$. Therefore we expect the optical Aharonov-Bohm effect 
in materials which do not have 
a curved spacetime counterpart satisfying the correspondence \eqref{ana1}, but do have a curl-free toroidal moment satisfying the analog condition of \eqref{g00}, namely
\begin{eqnarray}\label{g000}
\oint_C {\tau}_i dx^i \neq 0.
\end{eqnarray}
Taking this fact into account and also noting that design of metamaterials with simple  $\epsilon^{ij}$ or $\mu^{ij}$ is more feasible, 
in what follows we show the existence of optical Aharonov-Bohm effect in a (meta-)material with
isotropic $\epsilon^{ij} = \epsilon \delta^{ij}$ and $\mu^{ij} = \mu \delta^{ij}$  along with a toroidal moment satisfying \eqref{g000}.\\ 
In other words, as in the case of  the gravitational Aharonov-Bohm effect,
one expects a modified interference pattern for a coherent light ray which splits into two and then reunited to form a closed path on which  $\vec{{\tau}}$ is curl-free.
To this end we begin with the constitutive relations \eqref{D1} and \eqref{B1} in the presence of an antisymmetric magneto-electric 
coupling (i.e for ${\alpha}^{ij} = \frac{1}{4\pi} e^{ijk}\tau_{k}$), which could be written as follows
\begin{eqnarray}\label{g0}
\vec{B}=\mu\vec{H}+\vec{\tau}\times\vec{E} 
{\;\;\;\; , \;\;\;\;}\vec{D}=\varepsilon\vec{E}-\vec{\tau}\times\vec{H}
\end{eqnarray}
For the sake of simplicity, the medium  is taken to be isotropic, but $\epsilon$, $\mu$ and $\tau$ could be functions of spatial coordinates. Taking $\vec{E}$ and $\vec{H}$ as
\begin{eqnarray}\label{g02}
 \vec{E}({\vec{x}},t)=\vec{e}(\vec{x}) e^{ik[S(\vec{x})-ct]} \cr
 \vec{H}({\vec{x}},t)=\vec{h}(\vec{x}) e^{ik[S(\vec{x})-ct]} 
\end{eqnarray}
in which $S(\vec{x})$ is the spatial part of the  {\it eikonal}. Now taking the variations of both amplitudes $\vec{e}(\vec{x})$ and $\vec{h}(\vec{x})$ to be much 
smaller than the variation of the phase, namely $k S(\vec{x})$, and  
employing the geometrical optics approximation $k\rightarrow\infty$, Maxwell's equations translate into \cite{romer2006theoretical}
\begin{eqnarray}\label{g1}
\vec{{\nabla}}.\vec{B}=0{} \;\;\;\; \Rightarrow \;\;\;\; \mu\vec{\nabla}S.\vec{h}-\vec{\tau}.(\vec{\nabla}S\times\vec{e})=0
\end{eqnarray}
\begin{eqnarray}\label{g2}
\vec{{\nabla}}\times\vec{E}=-\frac{1}{c}\frac{{\partial}}{{\partial}t}
\vec{B} \;\;\;\; \Rightarrow \;\;\;\; (\vec{\nabla}S\times\vec{e})=\mu\vec{h}+\vec{\tau}\times\vec{e}
\end{eqnarray}
\begin{eqnarray}\label{g3}
\vec{{\nabla}}.\vec{D}=0{} \;\;\;\; \Rightarrow \;\;\;\; \varepsilon\vec{\nabla}S.\vec{e}+\vec{\tau}.(\vec{\nabla}S\times\vec{h})=0
\end{eqnarray}
\begin{eqnarray}\label{g4}
\vec{{\nabla}}\times\vec{H}=\frac{1}{c}\frac{{\partial}}{{\partial}t}
\vec{D}{} \;\;\;\; \Rightarrow \;\;\;\; (\vec{\nabla}S\times\vec{h})=-\varepsilon\vec{e}+\vec{\tau}\times\vec{h}
\end{eqnarray}
Equations \eqref{g2} and \eqref{g4} which imply that $\vec{h}$ and $\vec{e}$ are perpendicular could be rewritten in the following 
forms
\begin{eqnarray}\label{g51}
(\vec{\nabla}S-\vec{\tau})\times\vec{e}=\mu\vec{h}
\end{eqnarray}
\begin{eqnarray}\label{g52}
(\vec{\nabla}S-\vec{\tau})\times\vec{h}=-\varepsilon\vec{e},
\end{eqnarray}
implying that $(\vec{\nabla}S-\vec{\tau})$, $\vec{e}$, and $\vec{h}$ are mutually orthogonal. Combination of the last two equations yields
$(\vec{\nabla}S-\vec{\tau})\times[(\vec{\nabla}S-\vec{\tau})\times\vec{e}]=-\varepsilon\mu\vec{e}$, which upon 
using the orthogonality of 
$(\vec{\nabla}S-\vec{\tau})$ and  $\vec{e}$, we end up with the following eikonal equation
\begin{eqnarray}\label{g6}
{(\vec{\nabla}S-\vec{\tau})}^{2}=\varepsilon(\vec{x})\mu(\vec{x}).
\end{eqnarray}
The above equation resembles the Hamilton-Jacobi equation for charged particles in an electromagnetic field, in which the energy flux ($\vec{E} \times \vec{H}$) is 
in the direction of $(\vec{\nabla}S-\vec{\tau})$ which plays the role of {\it physical momentum}. \\
In this correspondence $\vec{\tau}$ and $\varepsilon(\vec{x})\mu(\vec{x})$
are playing the roles of the vector potential and scalar potential respectively or more precisely those of $\frac{e}{c}\vec{A}$ and $e \phi(x)$.
On this basis, one expects a Lorentz-like force to be acting on a light ray due to the non-vanishing curl of $\vec{{\tau}}$. 
Now assume that a light ray travels in a region of an optical medium with $\vec{\nabla}\times\vec{{\tau}}=0$. If $S_0$ is the solution of equation \eqref{g6} in 
the absence of $\vec{\tau}$, the general solution (i.e in the presence of $\vec{\tau}$) in the same region is given by
\begin{eqnarray}\label{g62}
S(\vec{x})=S_0 + \int_{\cal O}^{\vec{x}}\vec{\tau}.d\vec{x^\prime}
\end{eqnarray}
Integral ${\int}\vec{\tau}.d\vec{r}$ over a closed path is not necessarily zero unless $\vec{\nabla}\times\vec{{\tau}}$ vanishes over all the area enclosed by the path.  
As a consequence, interference effects due to this extra term are expected when a  coherent light beam splits in two and forms a closed path enclosing the area in which 
 $\vec{\nabla}\times\vec{{\tau}}\neq 0$.  
In other words, while the light trajectory is not influenced by the Lorentz-like force since $\vec{\nabla}\times\vec{{\tau}}=0$ along the path, we still have the
interference effects due 
to the non-zero $\tau$ through its line integral ${\int}\vec{\tau}.d\vec{r}$. This is the optical analog of the Aharonov-Bohm effect. \\
Noting that $\vec{\nabla}S = {\vec n}$ is the 
local refractive index \cite{romer2006theoretical}, the above analogy could be more clearly illustrated for $|\vec{\tau}| \ll 1$ which is the case for natural materials. 
To this end we use the identity 
\begin{eqnarray}\label{g61}
\vec{\nabla}S^2 = (\vec{\nabla}S - \vec\tau)^2 + 2\vec\tau . \vec{\nabla}S -\tau^2,
\end{eqnarray}
to arrive at the following form for the refractive index, 
\begin{eqnarray}\label{g6111} 
|\vec n| = |\vec{\nabla}S - \vec\tau| \left( 1 + \frac{2 \vec\tau . \vec{\nabla}S}{|\vec{\nabla}S - \tau|^2} - 
\frac{\tau^2}{|\vec{\nabla}S - \tau|^2}\right)^ {1/2} \;\;\;\;\; \cr
\;\;\;\;\;\;\;\; = |\vec{\nabla}S - \vec\tau| +\frac{ \vec\tau . (\vec{\nabla}S- \vec\tau)}{|\vec{\nabla}S - \vec\tau|} + {\cal O} (\tau^2).
\end{eqnarray}
Now from \eqref{g6} with $n_0^{2}=\varepsilon(\vec{x})\mu(\vec{x})$, the above relation reads   
\begin{eqnarray}\label{g611}
|\vec{n}| \simeq |{{n_{0}}}| + \vec{\tau}.{\hat{k}} 
\end{eqnarray}
where we have taken ${\hat k} = \frac{(\vec{\nabla}S- \vec\tau)}{|\vec{\nabla}S - \vec\tau|}$ because the vector $(\vec{\nabla}S- \vec\tau)$ is proportional to the 
Poynting vector.
This clearly has the same form  as \eqref{IRS2} with $\vec{\tau}$ playing the role of a vector potential, as expected from the established analogy.
Indeed for light rays with parallel and anti-parallel propagation directions to the toroidal moment, the above relation leads to 
\begin{eqnarray}\label{g65}
{|\vec{n}|}_{+} - {|\vec{n}|}_{-} \propto \vec{\tau}.\hat{k} 
\end{eqnarray}
a result in agreement with the non-reciprocal refraction which was shown to happen in a toroidal domain wall in multiferroic materials \cite{sawada2005optical}.
%%%%%%%%%%%%%%%%%%%%%%%%%%%%%%%%%%%%%%%%%%%%%%%%%%5
\section*{Summary and Discussion}
It would be useful to summarize our motivation and the follow up prescription, 
which led us to the prediction of an optical Aharonov-Bohm effect in toroidal (meta-)materials. We started with considering the gravitational analog 
of the usual (electromagnetic) Aharonov-Bohm effect in the context of threading formulation of spacetime decomposition. We then employed the 
analogy established between a curved 
background and a (meta-)material medium, or more specifically between a stationary spacetime's gravitomagnetic potential and the toroidal 
moment of its analog medium, to  predict the existence of an optical analog of the Aharonov-Bohm effect in toroidal (meta-)materials.\\
Indeed formally similar constitutive equations could be obtained for electromagnetic fields in a moving \cite{Landau2} or rotating medium \cite{vanblad}, 
in which the linear velocity of that medium or its 
angular velocity plays the role of the magnetic potential in magnetoelectric constitutive relations, and hence one could expect different analogs of the Aharonov-Bohm effect.
Considering the  classical wave equation for light propagation 
in a moving medium, it was shown that there exists a classical optical analog of Aharonov-Bohm effect in which velocity of the medium and its vortex play the roles 
of magnetic potential and magnetic field in the usual Aharonov-Bohm effect respectively \cite{Cook}. 
It is worth recalling that the more faithful gravitational analog of the Aharonov-Bohm effect arises in stationary spacetimes whose metric mixes 
space and time through which natural definitions of a gravitomagnetic potential and the corresponding gravitomagnetic field emerge.
On the other hand if we are going to design a metamaterial medium to exhibit an optical analog effect through the so called transformation 
media \cite{Pendry,Schurig}, the employed transformation should mix space and time coordinates. 
Indeed it is shown \cite {Leon1} that the spacetime transformation: 
$cdt=cdt' + ad\phi' \; , \; dr =\frac{dr'}{n} \; , \; d\phi = d\phi' \; , \; dz=\frac{dz'}{n}$, with constants $n$ and $a$,
corresponds to a  moving fluid forming a vortex with velocity ${\bf u} = (0, u_\phi, 0)$ which, without bending light, produces a
phase change proportional to $\oint{\bf u}. d{\bf r}$ on light enclosing the vortex, leading to an optical Aharonov-Bohm effect \cite{Leon3}. Now if we wish to design a 
metamaterial with a toroidal moment to exhibit
an optical Aharonov-Bohm effect, we should employ, as transformation media, the same generic coordinate transformation with the velocity of the vortex
replaced by the toroidal moment of the metamaterial.
Obviously one could read off the corresponding optical characteristics of the metamaterial from equations \eqref{ana1} and \eqref{ana2} using the metric 
tensor constructed from the above transformed spacetime differential elements.\\
Finally, a simple experimental setup for detecting this effect will probably include a coherent light beam splitting in two, each passing through the same 
medium with the same depth, permeability, and permittivity but with opposite rotation-free toroidal moments, before interfering again.\\
As a final note we should point out that our derivation of the optical AB effect is not only inspired by GR and its mathematical machinery,  but also due to 
the stablished opto-geometric duality, 
is not limited to the small values of the toroidal moment. Also our approach, compared to previous studies, establishes a more general and unifying picture of 
the different versions of the Aharonov-Bohm effect, including electromagnetic, gravitational and optical analogs of the effect.
%%%%%%%%%%%%%%%%%%%%%%%%%%%%%%%%%%%%%%%%%%%%%%%%%%%%%%%%%%%%%%%%%%%%%%%%%%%%
\section *{Acknowledgments}
The authors would like to thank the University of Tehran for supporting this project under the grants provided by the research council. M. N.-Z also 
thanks the high energy, cosmology, and astroparticle group at the Abdus Salam ICTP for kind hospitality during his visit when part of this study was 
carried out.
%%%%%%%%%%%%%%%%%%%%%%%%%%%%%%%%%%%%%%%%%%%%%%%%%%%%%%%%%%%%%%%%%%%
\textbf{\textit{}}

\end{document}